\documentclass[prl,twocolumn,showpacs,amsmath,amssymb,superscriptaddress,groupedaddress]{revtex4-1}

\usepackage{graphicx}
\usepackage{enumitem}
\setlist[enumerate]{label=\roman*)}
\usepackage{hyperref}
\hypersetup{
 pdfnewwindow=true, colorlinks=true,
 linkcolor=blue, anchorcolor=blue,
 citecolor=blue, filecolor=blue,
 menucolor=blue, urlcolor=blue}
\usepackage{breakurl}
\usepackage{dcolumn}

\usepackage{pdfpages}
\usepackage{pgffor}
\makeatletter
\AtBeginDocument{\let\LS@rot\@undefined}
\makeatother

\begin{document}

\title{Antiferromagnetism enables electron-phonon coupling in iron-based superconductors}

\author{Sinisa Coh}
\email{sinisacoh@gmail.com} 
\author{Marvin L. Cohen}
\author{Steven G. Louie}
\affiliation{Department of Physics, University of California at
  Berkeley and Materials Sciences Division, Lawrence Berkeley National
  Laboratory, Berkeley, California 94720, USA}

\date{\today}


\begin{abstract} 
It is shown that a generic form of an antiferromagnetic wave function opens strong electron-phonon coupling channels in the iron-based superconductors.  In the nonmagnetic state these channels exist locally on a single iron atom, but are canceled out between the two iron atoms in the primitive unit cell.  Our findings are based on symmetry and the presence of an $xz$/$yz$ Fermi surface near the M point and thus should be relevant for the known iron-based pnictide or chalcogenide superconductors.
\end{abstract}

\maketitle

There is much evidence that the superconducting state occurs in parallel with the antiferromagnetic state in iron-based superconductors.  Initially it was suggested in \cite{mazin2008} that fluctuations of the antiferromagnetic state induce electron pairing, and these electrons condense in the superconducting state.  Unlike conventional electron-phonon induced pairing, this interaction is repulsive so it requires the superconducting gap to change sign on the Fermi surface.  However, experiments on the latest generation \cite{wang2012,ge2014} of iron-based materials find a single pocket of carriers with an anisotropic but nodeless superconducting gap. \cite{PhysRevLett.112.107001,zhang2015superconducting} While this finding does not rule out all unconventional pairing symmetries, it is consistent with a conventional electron-phonon pairing.  In addition, scanning tunneling microscopy features found \cite{tang2015interface} in the FeSe monolayer above the superconducting gap are consistent with the calculated phonon spectral function \cite{coh2015} as well as with the kinks in the angle resolved photoemission spectra on a related material \cite{Malaeb2014}. Therefore, it is possible that the electron-phonon interaction can not be ignored in iron-based superconductors, and that early \cite{boeri2008,boeri2010} theoretical calculations may have underestimated its strength, as suggested in more recent calculations \cite{mangal2014,coh2015}.

If the electron-phonon interaction is important for pairing, it begs the obvious question: Is the parallel occurrence of superconductivity and antiferromagnetism just a coincidence?  We explore this question here.

\begin{figure}[t]
  \centering
  \includegraphics[width=3.4in]{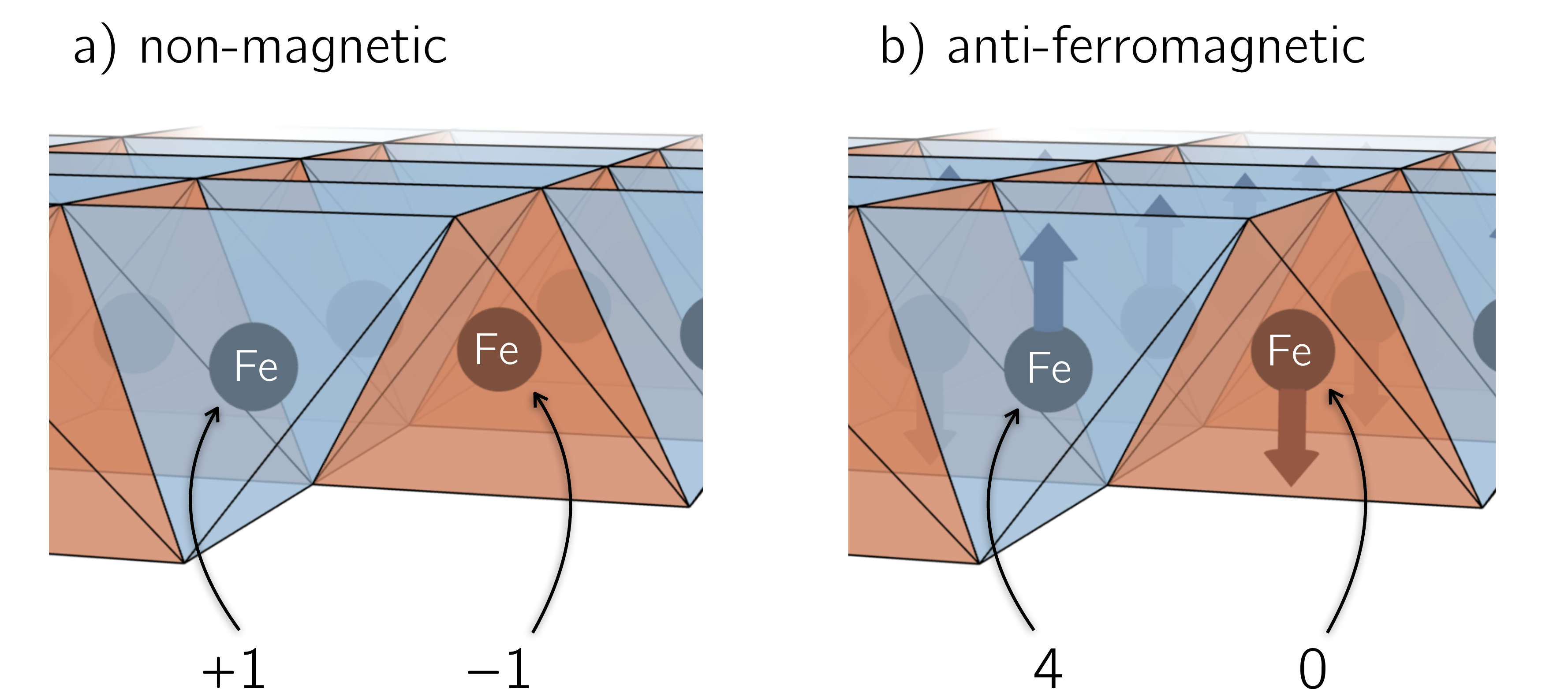}
  \caption{The contributions of the two iron atoms to the electron-phonon matrix element $g$ cancel each other in the nonmagnetic state, but not in the antiferromagnetic state.}
  \label{fig:structure}
\end{figure}

As shown below, the electron-phonon matrix element $g$ in these materials can be decomposed into contributions from individual atoms in the unit cell.  Representing with $\pm1$ the contribution of a single iron atom to $g$, we find in the nonmagnetic state,
$$g \simeq 1 - 1 = 0$$
as the two iron atoms in the cell contribute to $g$ with an opposing sign (see Fig.~\ref{fig:structure}). However, in an antiferromagnetic state, the iron $d$-like wave function $\Phi$ of a state of specific spin orientation is localized on only one of the two iron atoms in the cell.  Therefore only one iron atom per cell contributes to $g$ of this electronic state and the cancellation is prevented, resulting in
$$g \simeq 4 - 0 = 4.$$
The four-fold increase in the contribution of the active iron atom to $g$ is another consequence of the weight transfer in $\Phi$, and is discussed later. 

Our discussion here is mostly based on symmetry, but we rely on an earlier first-principles calculation \cite{coh2015} to obtain quantities that can not be inferred from symmetry alone (for example the orbital character of states near the Fermi level).  Our earlier first-principles calculations were done for an FeSe monolayer on a SrTiO$_3$ substrate and were focused on phonons in FeSe, not in the SrTiO$_3$ substrate.  Therefore, our results likely extend to most, if not all, iron-based superconductors, as they all contain layers of FeSe (or FeAs). In fact, several earlier studies \cite{boeri2010,bazhirov2012} on bulk materials found that electron-phonon matrix elements calculated in the antiferromagnetic state of the iron-based superconductors are larger than those in the nonmagnetic state (see also Fig.~S.1 in the Supplement Material \cite{supp}).  However, the microscopic origin of this increase remained unclear.

The focus of our paper is on the interplay between antiferromagnetism and electron-phonon interaction, not on the pairing symmetry. In general one expects that pairing may have contributions not only from electron-phonon interactions but also from other interactions such as magnetic or orbital fluctuations.\cite{PhysRevB.88.184505,PhysRevLett.112.027002}  Therefore, while our finding on electron-phonon interactions is general, it does not specifically address the resulting pairing symmetries across the families of iron-based superconductors.

We would also like to point out that the arguments presented in this paper assume that the xz/yz bands cross the Fermi level near the M point of the Brillouin zone.  Nevertheless, as far as we are aware, xz/yz bands cross the Fermi level near the M point in all the known iron-based pnictide or chalcogenide superconductors.

In what follows we first discuss two specific electronic states of an FeSe monolayer with the smallest primitive unit cell: the nonmagnetic (NM) and the checkerboard antiferromagnetic (cAFM) state.  Later we generalize our findings to nearly any ordered or disordered static antiferromagnetic state (with the correlation length exceeding the unit cell dimension).  We leave for future work the role of dynamic effects of the antiferromagnetic order.

\begin{figure}[!h]
  \centering
  \includegraphics[width=3.4in]{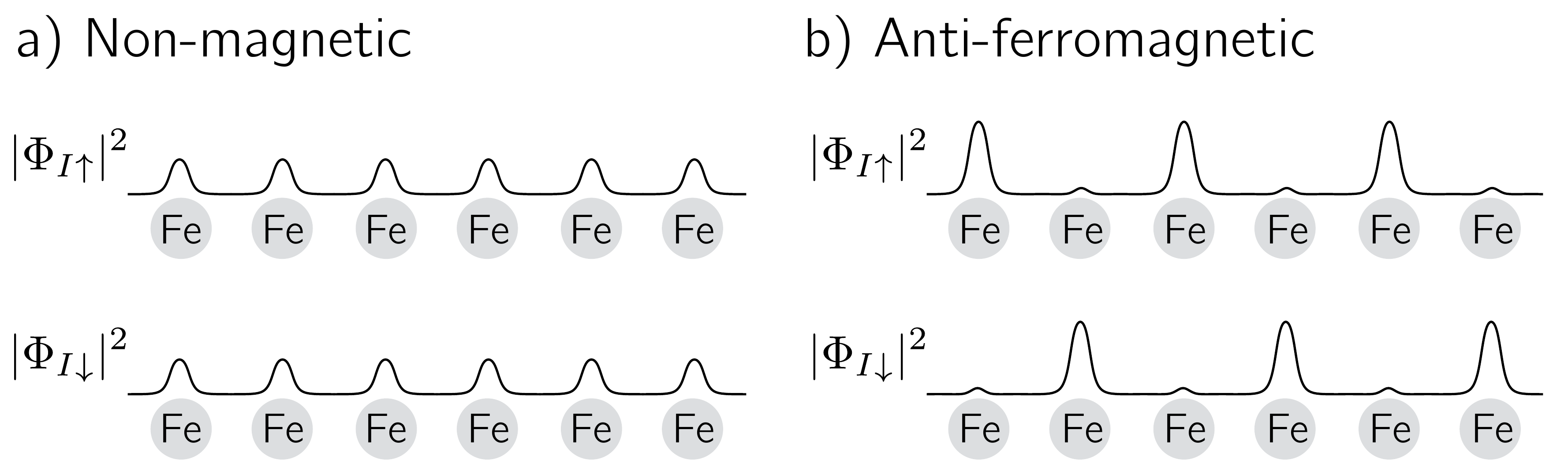}
  \caption{Sketch of the amplitude of electron orbital of up and down spin in the nonmagnetic (left) and the antiferromagnetic (right) state.  In our first-principles calculation for the antiferromagnetic state, the weight of the electron wave function crossing the Fermi level is 85\% on one of the iron atoms, and 5\% on the other iron atom, and the remaining weight is mostly on selenium $p$-states.}
  \label{fig:wfc}
\end{figure}

We work here within the density functional theory (DFT) framework where formally the exact ground state electron density $\rho$ is written in terms of the effective (Kohn-Sham) electron orbitals $\Phi_I$ that solve the Schrodinger-like equation with an effective one-body potential $V$.  The index $I$ on $\Phi_I$ refers to both the band index $n$ and the crystal momentum $\bf k$.  These orbitals characterize the quasiparticle states of the system.  In the spin-polarized variant of the theory orbitals for up and down spins might have different spatial dependences, $\Phi_{I \uparrow} ({\bf r}) \neq \Phi_{I \downarrow} ({\bf r})$. In particular, a (collinear) antiferromagnetic state has occupied electron orbitals for up and down spin located on a different subset of magnetic atoms in the unit cell.  For the simplest order, cAFM discussed earlier, the wave function $\Phi_{I \uparrow} ({\bf r})$ is mostly confined to one of the two iron atoms in the primitive unit cell while the corresponding $\Phi_{I \downarrow} ({\bf r})$ is on the other.  In the NM state, orbitals of both spin types exist at equal amplitudes on both Fe atoms in the cell, as sketched out in Fig.~\ref{fig:wfc}.

Even though the electron orbital $\Phi_I$ may be localized on only one of the Fe atoms in the unit cell,  with a well-defined crystal momentum it must periodically extend over all unit cells in the crystal.  Similarly, a phonon with a well defined momentum corresponds to the displacement of all atoms in the crystal.  Therefore as atoms vibrate around their equilibrium positions they change the effective potential $V \rightarrow V + \partial V_J$ across the entire crystal.  Here the phonon index $J$ replaces the phonon momentum $\bf q$ and the branch index $\nu$.  

The overlap integral of these three extended quantities,
$$g=\langle \Phi_I | \partial V_J | \Phi_{I'} \rangle ,$$
defines the electron-phonon matrix elements, one of the main ingredients that determines the electron pairing strength within the BCS mechanism of superconductivity.  However, representing $g$ in terms of extended quantities is not convenient for understanding the microscopic reason for the magnitude of $g$.  Therefore we rewrite the calculated extended electronic state $\Phi_I$ as a sum of functions $\phi_i$ highly localized on a single atom in the crystal, $\Phi_I = \sum_i \phi_i$. Later we give an explicit expression for $\phi_i$.
 Similarly, we rewrite the calculated extended potential change $\partial V_J$ due to a phonon as the sum of potential changes arising from the movement of individual atoms in the crystal, $\partial V_J = \sum_j \delta v_j$.

Now, following Ref.~\onlinecite{giustino2007}, an extended matrix element $g$ is rewritten as the sum of products of the localized quantities $\phi_i$, $\partial v_j$, and $\phi_{i'}$,
\begin{align}
g
&=
\langle \Phi_I | \partial V_J | \Phi_{I'} \rangle
\label{eq:ext_to_loc1}
\\
&=
\left( \sum_{i} \langle \phi_i | \right)
\left( \sum_{j} \partial v_j \right)
\left( \sum_{i'} | \phi_{i'} \rangle \right).
\label{eq:ext_to_loc2}
\end{align}
Since all three components $\phi_i$, $\partial v_j$, and $\phi_{i'}$ may be constructed using the Wannier function concept so that they are exponentially localized in real space, they will generally contribute to $g$ only when they are close to each other.  In the case of a FeSe monolayer we find that dominant contributions for electronic states at the Fermi level have either $i$, $i'$, and $j$ associated with the same iron atom, or $i$ and $i'$ on the same iron atom and $j$ on the neighboring selenium atom (numerical values are given in the Supplemental Material \cite{supp}).
Since orbitals $i$ and $i'$ are associated with a single iron atom we can group them together so that $g$ is rewritten as the sum over all iron atoms $a$ in the crystal,
\begin{align}
g
\simeq
\sum_{\rm atoms}
g_a.
\label{eq:decomposition}
\end{align}

We now make the analysis more concrete by rewriting the extended electron wave function $\Phi_I=\Phi_{n {\bf k}}$ as a vector $C$ 
in an explicit localized atomic basis $| a m \rangle$, such as a maximally localized Wannier function \cite{PhysRevB.56.12847} with an orbital character $m$,
\begin{align}
| \Phi_I \rangle = 
\sum_{a m} C^I_{am} e^{i {\bf k} \cdot {\bf r}_a} | a m \rangle.
\label{eq:phiDefinition}
\end{align}
Unless specified otherwise, the electron momentum $\bf k$ is defined in the two-iron atom unit cell.  Iron $d$-like orbital characters indexed with $m$ are always defined in the Cartesian frame of the one-iron atom cell (the $z$ axis is perpendicular to the FeSe plane).  Vector ${\bf r}_a$ points to atom $a$ and coefficients $C^I_{am}$ are cell-periodic by Bloch's theorem.  

Similarly, we rewrite $\partial V_J=\partial V_{\nu {\bf q}}$ in terms of a potential change $\partial v_{b \beta}$ due to the displacement of a single atom $b$ in the Cartesian direction $\beta$,
$$
\partial V_J = 
\sum_{b \beta} \xi^J_{b \beta} e^{i {\bf q} \cdot {\bf r}_b} \, \partial v_{b \beta}.
$$
Here $\xi$ is the polarization vector for phonon $J$ specifying the cell-periodic displacement of atom $b$ in direction $\beta$.

Inserting a decomposition of $\Phi$ and $\delta V$ into Eq.~\eqref{eq:ext_to_loc1} and using the simplification from Eq.~\eqref{eq:decomposition} after some algebra
we obtain the contribution of a single iron atom $a$ to $g$,
\begin{align}
g_a = 
\sum_{m m'} 
C^{I*}_{am} 
C^{I'}_{am'} 
\sum_{b \beta} 
\xi^J_{b \beta} 
e^{i {\bf q} \cdot \left( {\bf r}_b - {\bf r}_a \right) } 
\langle a m |
\partial v_{b \beta}
| a m' \rangle
.
\label{eq:ga_basis}
\end{align}
It is clear from the cell-periodicity of $C$ that $g_a$ is cell periodic as well, so it will be sufficient to compute $g_a$ only in the primitive unit cell.

In our previous report \cite{coh2015} we computed $g$ from first principles in a cAFM state of a FeSe monolayer and identified two channels, called 1 and 2, that have by far the largest $g$ among all states $I$ and $I'$ on the Fermi surface and among all phonon modes $J$.  Therefore in what follows we focus our discussion only on these two channels.  

Now we are ready to compute the sign of $g_a$ on both iron atoms in the primitive unit cell, for both channels (1 and 2).  The sign of $g_a$ is determined by the signs of the $C$'s, $\xi$, the exponential factor, and $\langle a m | \partial v_{b \beta} | a m' \rangle$ appearing in Eq.~\eqref{eq:ga_basis}.  We start by analyzing channel 2 in the NM state.

\begin{figure}
  \centering
  \includegraphics[width=3.4in]{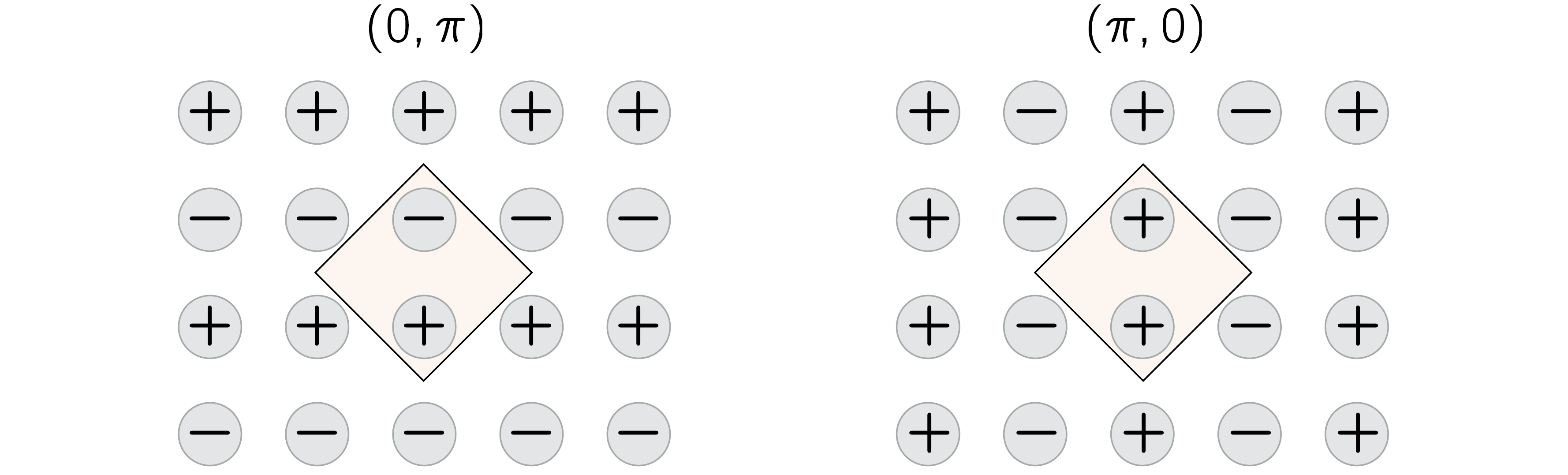}
 \caption{Signs of the electron wave function on the iron atoms (gray circles) at two edges of the one-iron unit cell.  The doubled unit cells are shown within the squares.}
 \label{fig:pipi}
\end{figure}

\begin{figure}
  \centering
  \includegraphics[width=3.4in]{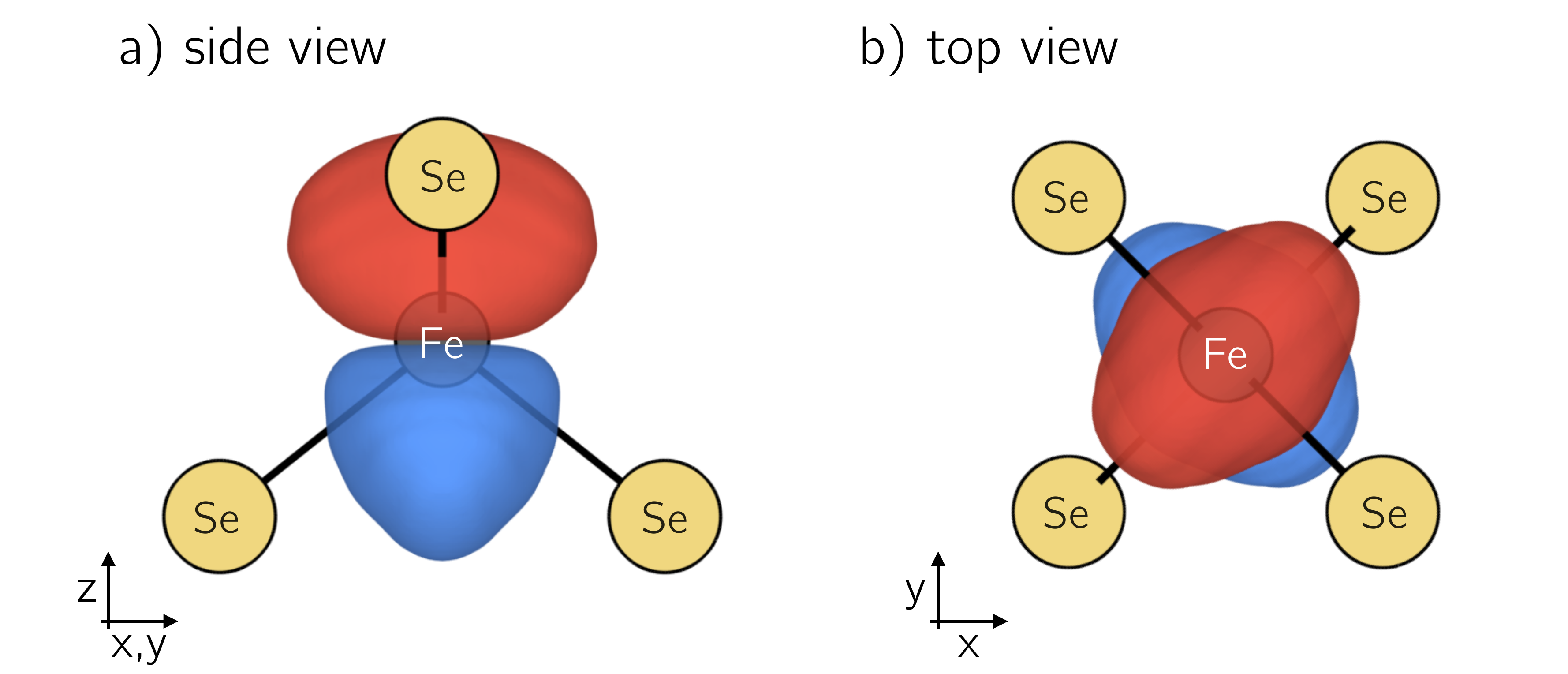}
  \caption{Isosurfaces of the potential induced by an upward displacement (along $z$) of the iron atom inside the tetrahedral selenium cage.  The ellipsoidal asymmetry in the iso-surfaces along the $x$ and $y$ axes indicate the presence of a $d$-like $xy$ component of the induced potential. Cartesian labels follow the one-iron unit cell convention.  The red (blue) isosurface is drawn at a constant value of the induced potential. This constant value is set at $2$\% of the maximal (minimal) value of the induced potential.}
  \label{fig:dxy}
\end{figure}

\begin{enumerate}
\item 
Channel 2 scatters states originating from $d$-like $xz$ states to those of $yz$ character, and vice versa.  In the two-iron unit cell both $xz$ and $yz$ states cross the Fermi level near the ${\bf k}=(\pi,\pi)$ point, but in the unfolded one-iron unit cell one state is near $(0,\pi)$ and the other near $(\pi,0)$.  Therefore, as illustrated in Fig.~\ref{fig:pipi}, in the two-iron unit cell $\Phi$ either has opposite sign on the two iron atoms in the cell, or has the same sign.  Which is which depends on the choice of the basis atoms in the two-iron cell, and is not relevant for the following discussion.  Recalling the definition of the coefficients $C$ from Eq.~\eqref{eq:phiDefinition} and using ${\bf k} = (\pi,\pi)$ we conclude that the relative sign of $C$ is the same on both iron atoms for one of the states and the opposite for the other state.   

\item 
The phonon eigenvector $\xi$ in channel 2 has the opposite sign on the two iron atoms in the unit cell, as it consists of an out of phase vertical displacement of iron atoms, with first neighboring iron atoms moving in opposite directions.  In the one-iron unit cell this mode would have ${\bf q}=(\pi,\pi)$, consistent with the fact that this channel scatters state ${\bf k}=(0,\pi)$ to state $(\pi,0)$.

\item The exponential factor $e^{i {\bf q} \cdot \left( {\bf r}_b - {\bf r}_a \right) }$ equals $1$ as a vertical displacement of an iron atom dominantly affects the localized orbital on the same iron atom. Therefore $a=b$, so ${\bf r}_a={\bf r}_b$, and $e^{i {\bf q} \cdot \left( {\bf r}_b - {\bf r}_a \right) }=1$.

\item
Finally, since channel 2 couples a $d$-like $xz$ to $yz$ state, the matrix element $\langle a m | \partial v_{b \beta} | a m' \rangle$ is dominated by the induced potential $\partial v_{b \beta}$ with $xy$ character (since $\langle xz| xy | yz \rangle \neq 0$).  By symmetry, displacement of the iron atom perpendicular to the FeSe plane can create this kind of potential only through an interaction with neighboring selenium atoms, as shown in Fig.~\ref{fig:dxy} and discussed later in more detail.  However, selenium tetrahedra formed around the two iron atoms in the unit cell are inverted images of each other (compare the red and blue tetrahedra in Fig.~\ref{fig:structure}) so the induced $xy$ potential has an opposite sign as well, as confirmed by the explicit calculation of the matrix element given in the Supplemental Material \cite{supp}.

\end{enumerate}

Since the relative sign of the two iron atoms is opposite for an odd number of factors (i, ii, and iv) we conclude that the relative sign of $g_a$ is opposite as well.  Therefore, the electron-phonon matrix element $g$ for channel 2 vanishes in the nonmagnetic state.

In the cAFM case the wave function coefficient $C$ for a state at the Fermi level is zero for one of the Fe atoms in the unit cell (see Fig.~\ref{fig:wfc}), so the corresponding $g_a$ is zero as well.  This prevents cancellation of the contributions between two atoms in the unit cell.  The remaining iron atom has $g_a$ four times larger than in the NM case.  The reason for this increase is the two-fold increase in the specific spin density of the iron atom in the cAFM state compared to the NM state.  Therefore the square of the wave function coefficient $|C|^2$ appearing in Eq.~\eqref{eq:ga_basis} is increased by a factor of two.  Another factor of two originates from the twofold increase in the $xy$-like induced potential $\delta v_{b\beta}$ upon vertical displacement of an iron atom (see also \cite{supp} for numerical values). 

\begin{figure}
  \centering
  \includegraphics[width=3.4in]{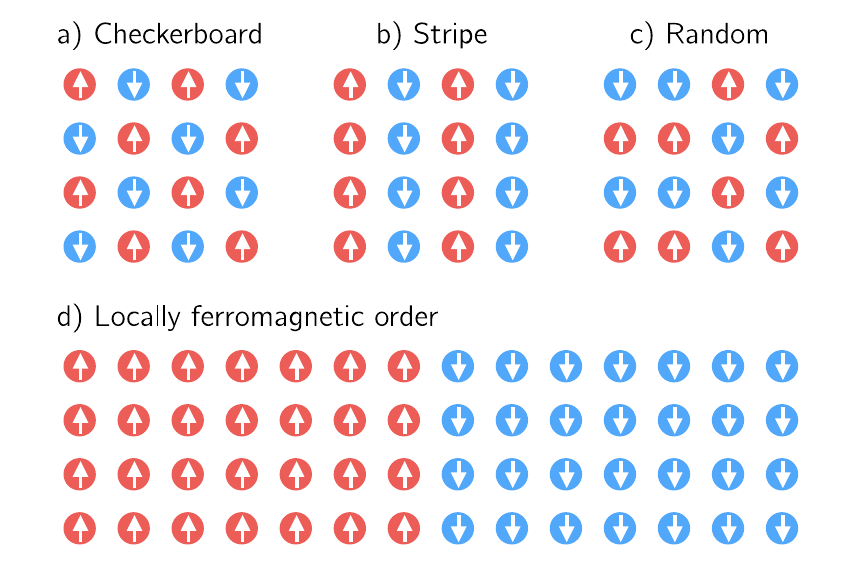}
  \caption{Four examples of antiferromagnetic order for iron atoms (red and blue circles denote iron atoms with opposite spin).  In the first three panels (a, b, and c) antiferromagnetic order enables electron-phonon interaction.  However, the order shown in panel d doesn't enable electron-phonon interaction since most first-neighboring spins are ferromagnetically arranged.}
  \label{fig:order}
\end{figure}

Generalization to nearly any kind of antiferromagnetic order is now straightforward. 
Let us consider a large $N \times N$ supercell of FeSe with an arbitrary antiferromagnetic ordering of spins (as shown in panels a, b, and c of Fig.~\ref{fig:order}).  For any such order, we can construct a pattern of atom displacements in which all up-spin iron atoms move vertically above the FeSe plane, and all down-spins move into the plane (or vice-versa).  One can easily check from our previous analysis that the electron-phonon matrix element $g$ for this displacement pattern will be as large as in the cAFM state.  For example, it would be enough to show that, starting from the cAFM order in the supercell, exchanging any pair of opposing spins will not affect $g$ as long as the direction of the atomic displacements $\xi$ of the same pair is exchanged as well.  

This argument would not apply to the case, although formally antiferromagnetically ordered, where most neighboring spins are locally arranged ferromagnetically.  One such example of formally antiferromagnetic, but actually ferromagnetic order, is shown in panel d of Fig.~\ref{fig:order}.  In this scenario $g$ must be nearly zero as the displacement pattern given by our construction would correspond to a rigid translation of all atoms on the same side of the cell. (Here we can safely ignore vertical displacement of selenium atoms as we find them to have a negligible contribution to $g$.)  This result enforced by translation symmetry can be obtained from Eqs.~\ref{eq:ext_to_loc2} and \ref{eq:decomposition} by including the off-site matrix element where $i$ and $i'$ are on the first-neighboring iron atoms.

The discussion of channel 1 reaches a conclusion similar to that of channel 2:  in the NM state, the two iron atoms in the unit cell contribute to $g$ with opposite signs.  Phonon coupling in channel 1 is a soft mode responsible for condensation of the so called orthorhombic (nematic) ground state in bulk FeSe \cite{mcqueen2009}.  Therefore this mode consists of an in-plane displacement of both iron and selenium atoms.  However, only the in-plane displacement of selenium atoms contributes to $g_a$.  In particular, coupling is large only when selenium atom $b$ moves towards or away from the neighboring iron atom $a$ (see Supplemental Material \cite{supp} for explicit numerical values of $g_a$).  Since $a\neq b$ we have ${\bf r}_a \neq {\bf r}_b$ and the exponential term appearing in Eq.~\eqref{eq:ga_basis} leads to the dependence of matrix element $g$ on the phonon momentum $\bf q$.  We find that $g\sim |{\bf q}|$ for small $\bf q$ and therefore conclude that the forward scattering is greatly suppressed over the backward scattering in channel 1. Another difference with respect to channel 2 is that states coupled in channel 1 have the same orbital character (the $d$-like $xz$ state couples to $xz$, and $yz$ to $yz$). 

Our findings rely on having xz/yz bands crossing the Fermi level near the M point as found experimentally.  However, some DFT calculations result in an electronic structure that is inconsistent with experiment. In these cases, our mechanism may or may not apply. One such example is the DFT calculated band structure for the striped phase. Nevertheless, the calculated electron-phonon interaction with the incorrect band structure is also stronger than in the nonmagnetic case. Sorting out these theoretical observations is worthy of a future study, but it is beyond the focus and scope of our current work.

In closing, we discuss the role of the selenium height on the magnitude of the electron-phonon matrix element $g$.   In channel 1 the dominant contribution to $g$ comes from the displacement of selenium atoms, so it is not surprising that the position of selenium atom is relevant for $g$ in that channel.  Somewhat less expected is our finding that the position of a selenium atom is crucial for coupling in channel 2, since it involves displacement of iron atoms.  As mentioned earlier, this channel scatters a $d$-like $xz$ state to $yz$ and therefore the potential induced by the upward movement of the iron atom must have a $xy$-like component.  However, iron atoms are all in the same plane, so the $xy$-like potential component must originate from the interaction of an iron atom with neighboring tetrahedrally bonded selenium atoms.  We confirmed this finding with an explicit first-principles calculation on a $3 \times 3$ FeSe unit cell. Upward displacement of a single iron atom in this enlarged cell transfers the electron's charge into the Fe-Se bond above the iron plane (making the bond more covalent), and out of the Fe-Se bond below the plane.  Therefore, the induced potential has some $xy$ component as shown in Fig.~\ref{fig:dxy}.

This means that $g$ in channel 2 is maximized when the direction of the iron-selenium bond is aligned with the extremal points of product $\Phi_{I}^{\star} \Phi_{I'}$.  Since the electron wave functions have $d$-like $xz$ and $yz$ character, $g$ is maximal when Fe-Se bonds form an ideal tetrahedron.  This is consistent with the empirical finding that an ideal tetrahedral environment gives the highest superconducting transition temperature \cite{chul2008}.

\begin{acknowledgments}
This work was supported by the National Science Foundation under Grant No. DMR-1508412 which provided for the theoretical analysis and the Theory of Materials Program at the Lawrence Berkeley National Lab through the Office of Basic Energy Sciences, U.S. Department of Energy under Contract No. DE-AC02-05CH11231 which provided for the electron-phonon coupling calculations.  This research used resources of the National Energy Research Scientific Computing Center, which is supported by the Office of Science of the US Department of Energy.
\end{acknowledgments}

\bibliography{pap}

\foreach \x in {1,2,3}
{%
\clearpage


\section*{Supplementary material}

# Supplemental material for "Anti-ferromagnetism enables electron-phonon coupling in iron-based superconductors"

Sinisa Coh, Marvin L. Cohen, and Steven G. Louie

*Department of Physics, University of California at Berkeley and Materials Sciences Division,*
*Lawrence Berkeley National Laboratory, Berkeley, California 94720, USA*

(Dated: December 17, 2015)

This supplement contains the calculated values of the following electron-phonon matrix elements, $\langle am|\partial v_{b\beta}|a'm'\rangle$ used in the main text. Here bra and ket are localized orbitals on one of the two iron atoms in the unit cell with $d$-like $xz$ or $yz$ orbital character. Object $\partial v_{b\beta}$ is the potential induced by a displacement of a single atom in the crystal (without moving its periodic images). For completeness we will consider displacement of all four atoms (two Fe and two Se) in the unit cell ($b$) in all three Cartesian directions ($\beta$).

Here are coordinates of four atoms (two Fe and two Se) in the FeSe monolayer unit cell in reduced coordinates (for definiteness selenium height parameter $z$ is larger than zero),

|      |     |     |     |
|------|-----|-----|-----|
| Fe   | 0.5 | 0   | 0   |
| Fe′  | 0   | 0.5 | 0   |
| Se   | 0   | 0   | -z  |
| Se′  | 0.5 | 0.5 | z   |

By convention, $m$ index (i.e. $xz$ or $yz$ orbital) is given in the reduced unit cell with one iron atom per cell, as this is a prevailing choice in the literature. However, $\beta$ index (atom displacement direction) here is given in the crystallographic unit cell containing two-iron atoms per cell. The one-iron cell is rotated by $45°$ with respect to the two-iron cell so that its $x$ axis points along the (110) direction in the two-iron cell.

We will list the electron-phonon matrix element as a $4 \times 4$ matrix in $a$, $m$, $a'$, and $m'$ indices for a fixed choice of atom displacement ($b$, $\beta$) with the following format,

$$\begin{pmatrix} \langle\text{Fe}:xz|\partial v|\text{Fe}:xz\rangle & \langle\text{Fe}:xz|\partial v|\text{Fe}:yz\rangle & \langle\text{Fe}:xz|\partial v|\text{Fe}':xz\rangle & \langle\text{Fe}:xz|\partial v|\text{Fe}':yz\rangle \\ \langle\text{Fe}:yz|\partial v|\text{Fe}:xz\rangle & \langle\text{Fe}:yz|\partial v|\text{Fe}:yz\rangle & \langle\text{Fe}:yz|\partial v|\text{Fe}':xz\rangle & \langle\text{Fe}:yz|\partial v|\text{Fe}':yz\rangle \\ \langle\text{Fe}':xz|\partial v|\text{Fe}:xz\rangle & \langle\text{Fe}':xz|\partial v|\text{Fe}:yz\rangle & \langle\text{Fe}':xz|\partial v|\text{Fe}':xz\rangle & \langle\text{Fe}':xz|\partial v|\text{Fe}':yz\rangle \\ \langle\text{Fe}':yz|\partial v|\text{Fe}:xz\rangle & \langle\text{Fe}':yz|\partial v|\text{Fe}:yz\rangle & \langle\text{Fe}':yz|\partial v|\text{Fe}':xz\rangle & \langle\text{Fe}':yz|\partial v|\text{Fe}':yz\rangle \end{pmatrix}.$$

The electron-phonon matrix elements below are given in atomic (Rydberg) units. For a more convenient representation we multiplied all elements by $10^3$ and rounded up to zero all elements less than $0.5 \cdot 10^{-3}$. Dominant terms corresponding to channels 1 and 2 discussed in the main text are highlighted in bold.

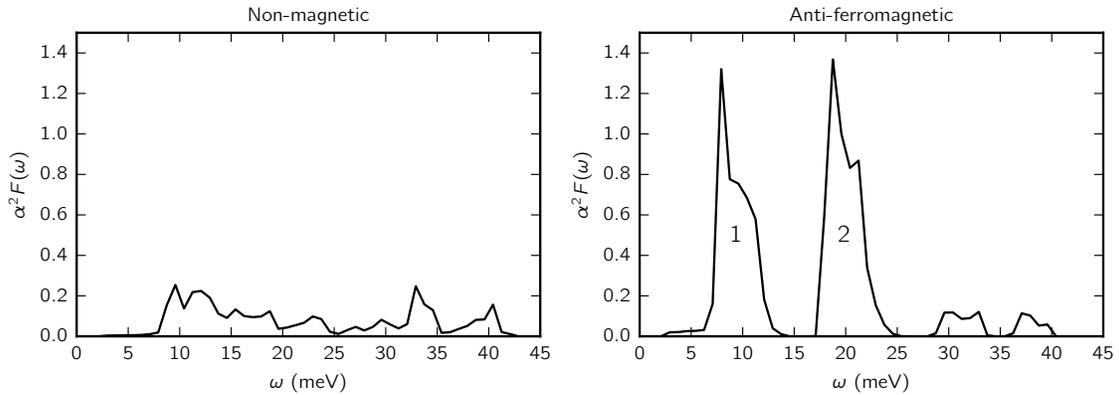

FIG. S.1. The integrated electron-phonon coupling $\alpha^2 F(\omega)$ in the NM and cAFM state of FeSe monolayer within the GGA+A approach. Clearly, $\alpha^2 F(\omega)$ is larger in the cAFM case than in the NM case. This qualitative result does not depend on the value of parameter $A$.

Orbitals $\langle am|$ and $|a'm'\rangle$ below have a well defined (and equal) spin direction. In the cAFM case value of the matrix element depends on the spin of the orbital. The numerical values below are given for a spin component that is occupied on the Fe atom and empty on the Fe$'$ atom.

- Displacement of $b =$ Fe atom

  – nonmagnetic (NM) state

$$\begin{pmatrix} 0 & 0 & 8 & -6 \\ 0 & 0 & -6 & -19 \\ 8 & -6 & 4 & 0 \\ -6 & -19 & 0 & 6 \end{pmatrix}^{\beta=x} \begin{pmatrix} 0 & 0 & -8 & -6 \\ 0 & 0 & -6 & 18 \\ -8 & -6 & -4 & 0 \\ -6 & 18 & 0 & -6 \end{pmatrix}^{\beta=y} \begin{pmatrix} 0 & -\mathbf{45} & 0 & -4 \\ -\mathbf{45} & 0 & -12 & 0 \\ 0 & -12 & 0 & -3 \\ -4 & 0 & -3 & 0 \end{pmatrix}^{\beta=z}$$
(S.1)

  – antiferromagnetic (cAFM) state

$$\begin{pmatrix} -1 & 1 & 12 & -6 \\ 1 & -1 & -3 & -26 \\ 12 & -3 & -5 & 0 \\ -6 & -26 & 0 & 2 \end{pmatrix}^{\beta=x} \begin{pmatrix} 0 & 0 & -9 & -5 \\ 0 & -1 & -3 & 22 \\ -9 & -3 & 5 & 0 \\ -5 & 22 & 0 & -2 \end{pmatrix}^{\beta=y} \begin{pmatrix} 0 & -\mathbf{100} & 0 & -5 \\ -\mathbf{100} & 0 & -8 & 0 \\ 0 & -8 & 0 & 8 \\ -5 & 0 & 8 & 0 \end{pmatrix}^{\beta=z}$$
(S.2)

- Displacement of $b =$ Fe$'$ atom

  – nonmagnetic (NM) state

$$\begin{pmatrix} -4 & 0 & -8 & 5 \\ 0 & -6 & 6 & 18 \\ -8 & 6 & 0 & 0 \\ 5 & 18 & 0 & 0 \end{pmatrix}^{\beta=x} \begin{pmatrix} 4 & 0 & 8 & 5 \\ 0 & 6 & 6 & -19 \\ 8 & 6 & 0 & 0 \\ 5 & -19 & 0 & 0 \end{pmatrix}^{\beta=y} \begin{pmatrix} 0 & 3 & 0 & 12 \\ 3 & 0 & 4 & 0 \\ 0 & 4 & 0 & \mathbf{45} \\ 12 & 0 & \mathbf{45} & 0 \end{pmatrix}^{\beta=z}$$
(S.3)

  – antiferromagnetic (cAFM) state

$$\begin{pmatrix} -10 & 0 & -19 & 4 \\ 0 & -9 & 6 & 33 \\ -19 & 6 & 0 & 0 \\ 4 & 33 & 0 & 0 \end{pmatrix}^{\beta=x} \begin{pmatrix} 10 & 0 & 17 & 12 \\ 0 & 9 & 8 & -30 \\ 17 & 8 & 0 & 0 \\ 12 & -30 & 0 & 0 \end{pmatrix}^{\beta=y} \begin{pmatrix} 0 & 15 & 0 & 13 \\ 15 & 0 & 0 & 0 \\ 0 & 0 & 0 & 7 \\ 13 & 0 & 7 & 0 \end{pmatrix}^{\beta=z}$$
(S.4)

- Displacement of $b =$ Se atom

  – nonmagnetic (NM) state

$$\begin{pmatrix} -\mathbf{24} & 6 & 0 & 3 \\ 6 & -\mathbf{24} & 0 & 1 \\ 0 & 0 & 7 & 0 \\ 3 & 1 & 0 & -7 \end{pmatrix}^{\beta=x} \begin{pmatrix} 7 & 0 & 0 & 0 \\ 0 & -7 & -3 & 1 \\ 0 & -3 & -\mathbf{24} & -6 \\ 0 & 1 & -6 & -\mathbf{24} \end{pmatrix}^{\beta=y} \begin{pmatrix} -4 & 5 & -1 & 2 \\ 5 & -4 & -2 & 1 \\ -1 & -2 & -4 & -5 \\ 2 & 1 & -5 & -4 \end{pmatrix}^{\beta=z}$$
(S.5)

  – antiferromagnetic (cAFM) state

$$\begin{pmatrix} -\mathbf{42} & 14 & 1 & 2 \\ 14 & -\mathbf{42} & 0 & 1 \\ 1 & 0 & 2 & 0 \\ 2 & 1 & 0 & -2 \end{pmatrix}^{\beta=x} \begin{pmatrix} 13 & 0 & -1 & 0 \\ 0 & -13 & -2 & 2 \\ -1 & -2 & 4 & -1 \\ 0 & 2 & -1 & 4 \end{pmatrix}^{\beta=y} \begin{pmatrix} -24 & 11 & -1 & 1 \\ 11 & -24 & -2 & 1 \\ -1 & -2 & 33 & 0 \\ 1 & 1 & 0 & 33 \end{pmatrix}^{\beta=z}$$
(S.6)

- Displacement of $b = \text{Se}'$ atom

    - nonmagnetic (NM) state

$$\begin{pmatrix} -7 & 0 & 0 & 0 \\ 0 & 7 & -3 & -1 \\ 0 & -3 & \mathbf{24} & -6 \\ 0 & -1 & -6 & \mathbf{24} \end{pmatrix}^{\beta=x} \begin{pmatrix} \mathbf{24} & 6 & 0 & 3 \\ 6 & \mathbf{24} & 0 & -1 \\ 0 & 0 & -7 & 0 \\ 3 & -1 & 0 & 7 \end{pmatrix}^{\beta=y} \begin{pmatrix} 4 & 5 & 1 & 2 \\ 5 & 4 & -2 & -1 \\ 1 & -2 & 4 & -5 \\ 2 & -1 & -5 & 4 \end{pmatrix}^{\beta=z} \quad (\text{S.7})$$

    - antiferromagnetic (cAFM) state

$$\begin{pmatrix} -13 & 0 & 2 & -1 \\ 0 & 13 & -2 & -2 \\ 2 & -2 & -4 & -1 \\ -1 & -2 & -1 & -4 \end{pmatrix}^{\beta=x} \begin{pmatrix} \mathbf{42} & 14 & -1 & 2 \\ 14 & \mathbf{42} & 0 & -1 \\ -1 & 0 & -2 & 0 \\ 2 & -1 & 0 & 2 \end{pmatrix}^{\beta=y} \begin{pmatrix} 24 & 11 & 1 & 1 \\ 11 & 24 & -2 & -1 \\ 1 & -2 & -33 & 0 \\ 1 & -1 & 0 & -33 \end{pmatrix}^{\beta=z} \quad (\text{S.8})$$


}

\end{document}